\documentclass{WileyMSP-template}

\usepackage{graphicx}
\usepackage{tabularx}
\usepackage{multirow}
\usepackage{amsmath, amssymb, amsfonts, bm}

\begin{document}

\pagestyle{fancy}
\rhead{}

\title{Ultra-broadband Noise-Insulating Periodic Structures Made of Coupled Helmholtz Resonators}

\maketitle

\author{Mariia Krasikova*}
\author{Aleksandra Pavliuk}
\author{Sergey Krasikov}
\author{Mikhail Kuzmin}
\author{Andrey Lutovinov}
\author{Anton Melnikov}
\author{Yuri Baloshin}
\author{David A.~Powell}
\author{Steffen Marburg}
\author{Andrey Bogdanov}

\dedication{Mariia Krasikova and Aleksandra Pavliuk contributed equally to this work.}

\begin{affiliations}
M. Krasikova, A. Pavliuk, S. Krasikov, M. Kuzmin, A. Lutovinov, Y. Baloshin, A. Bogdanov\\
School of Physics and Engineering, ITMO University, Saint Petersburg 197101, Russia\\
Email Address: mariia.krasikova@metalab.ifmo.ru

M. Krasikova, A. Melnikov, S. Marburg\\
Chair of Vibroacoustics of Vehicles and Machines, Technical University of Munich, Garching b. München 85748, Germany

D. A. Powell\\
School of Engineering and Technology, University of New South Wales, Northcott Drive, Canberra, Australian Capital Territory 2600, Australia

A. Bogdanov\\
Qingdao Innovation and Development Center of Harbin Engineering University, Sansha road 1777, Qingdao 266404, China
\end{affiliations}


\begin{abstract}
Acoustic metamaterials and phononic crystals represent a promising platform for the development of noise-insulating systems characterized by a low weight and small thickness. Nevertheless, the operational spectral range of these structures is usually quite narrow, limiting their application as substitutions of conventional noise-insulating systems. In this work, the problem is tackled by demonstration of several ways for the improvement of noise-insulating properties of the periodic structures based on coupled Helmholtz resonators. It is shown that tuning of local coupling between the resonators leads to the formation of ultra-broad stop-bands in the transmission spectra. This property is linked to band structures of the equivalent infinitely periodic systems and is discussed in terms of band-gap engineering. The local coupling strength is varied via several means, including introduction of the so-called chirped structures and lossy resonators with porous inserts. The stop-band engineering procedure is supported by genetic algorithm optimization and the numerical calculations are verified by experimental measurements.
\end{abstract}


\section{Introduction}
One of the global challenges of the modern world is noise pollution, the level of which steadily increases every year~\cite{stansfeld2003noise,munzel2021transportation,thompson2022noise,berkhout2023anthropogenic}. Conventional means of limiting the spread of noise include passive noise-insulation systems with the efficiency depending on the mass and volume of the materials from which the system is made of~\cite{bies2017engineering}. An alternative approach arose from the field of acoustic metamaterials and phononic crystals demonstrating unprecedented ways of wave control and manipulation~\cite{ma2016acoustic,cummer2016controlling,assouar2018acoustic,vasileiadis2021progress,muhammad2022photonic}. Extraordinary properties of these structures allow not only to implement superlenses~\cite{kaina2015negative,ma2022acoustic}, invisibility cloaks~\cite{chen2007acoustic,chen2010acoustic}, or analog computing schemes~\cite{zangeneh-nejad2021analogue} but also various noise-mitigating structures~\cite{kumar2019present,romero2020design,gao2022acoustic} characterized by high efficiency as well as ventilation properties~\cite{kumar2020recent,ang2023systematic}. However, the practical application of such structures is still limited by a typically narrow frequency range of operation and complex geometry.
These problems were addressed via the proposal of new designs~\cite{gao2022acoustic,xiao2022multifunctional,ang2023systematic,trematerra2023noise,jin2023reconfigurable} and the implementation of various optimization algorithms including those based on machine learning~\cite{bacigalupo2020machinelearning,gurbuz2021generative,muhammad2022machine}.
Within the variety of approaches, one of the routes for the engineering of noise-insulating properties may be borrowed from the field of photonics in which the development of metamaterials has gone far ahead~\cite{zheludev2012metamaterials,kadic20193d,xiao2020active,solntsev2021metasurfaces}. For instance, it was demonstrated that bandgaps in photonic crystals can be broadened by the slow variation of the layers' thickness~\cite{baets1987high,saka1993bragg,wu2009analysis}. Structures in which parameters are gradually varied across the unit cells are usually called to be \textit{chirped}. The associated effect of band-gaps broadening was also observed in plasmonic structures~\cite{bouillard2012broadband} and even in nature where the effect is manifested as, for instance, metallic-like shine of some beetles’ shells~\cite{thompson2004vision,yu2013biomimetic,cook2016theory}.
Returning to the case of phononic crystals, chirped structures were studied for the development of acoustic diodes~\cite{cebrecos2016asymmetric}, wave concentrators~\cite{romero-garcia2013enhancement} and rainbow trapping systems~\cite{tian2017rainbow}. 

In this work, we assume that since chirping may result in the broadening of band-gaps the associated noise-insulating properties of acoustic structures may be enhanced as a result.
In order to stick to a particular design, we consider periodic structure based on strongly coupled Helmholtz resonators demonstrated in Ref.~\cite{krasikova2023metahouse}. Using the physics-based approach supported by genetic algorithm optimization we demonstrate that chirping of local coupling between the resonators may lead to the formation of broad stop-band in the transmission spectra, which was observed numerically and experimentally. These properties are associated with the overlap of band-gaps of structures with different unit cells. Moreover, we demonstrate that the results can be generalized and that the recipe for the stop-band engineering implies tuning of local coupling between the resonators. This concept was demonstrated not only for chirped structures, but also for the structures based on super-cells and lossy resonators with porous materials inserted into their body.

\begin{figure}[htbp!]
    \centering
    \includegraphics[width=\linewidth]{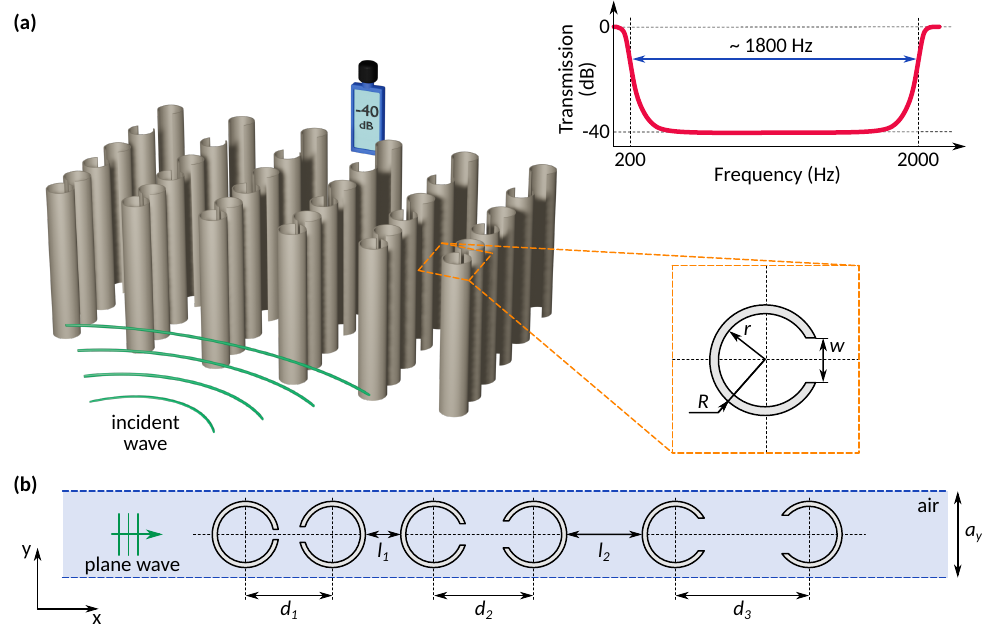}
    \caption{\textbf{Schematic illustration of the system under the consideration.} 
    (a) Artistic illustration of the considered periodic structures consisting of Helmholtz resonators. The optimized structures are capable to suppress noise up to $-40$~dB in a broad range of frequencies, about $200$--$2100$~Hz. Resonators have C-shaped are characterized by the outer radius $R$, the inner radius $r$ and the slit width $w$. Note that the actual distance between the rays of the resonators is smaller than in the illustation.
    (b) Semi-infinite structure, which is finite along the direction of wave propagation and infinitely periodic along other direction. Period in this case is $a_y$, distances within the pairs are $d_{1-3}$ and distances between the pairs are $l_{1,2}$.}
    \label{fig:system}
\end{figure}

\section{Results and Discussion}
\subsection{System description}
The considered structures are made of coupled Helmholtz resonators representing pipes with slits carved along their rotation axes as shown in \textbf{Figure~\ref{fig:system}}. For the sake of simplicity, it is assumed that the resonators are infinitely long meaning that effectively the system is two-dimensional. Normal modes of the similar C-shaped resonators as well as the associated scattering properties were extensively studied in Ref.~\cite{moheit2020analysis}. This type of resonators was also used for demonstration of the strong Willis coupling~\cite{melnikov2019acoustic} and development of the noise-mitigating capsule~\cite{melnikov2020acoustic}.
In this work the resonators have fixed geometric parameters, such that the outer radius is $R = 53$~mm, the inner radius is $r = 48$~mm.
All of the resulting spectra are compared with the initial structure with the slit width $w = 40$~mm and the distance between the centers of the resonators $120$~mm.
For the purpose of the work the material from which the resonators are made of is irrelevant and it can be any solid material with an impedance significantly larger than that of air (see Methods section). In particular, the samples for the experimental measurements were fabricated via 3D printing using polylactic acid filament. 

To form a periodic structure the resonators are arranged in pairs  as it allows to broaden bang-gaps due to the strong local coupling~\cite{krasikova2023metahouse}. The studies of transmission properties are performed for the semi-infinite structures which in this work are understood as structures having finite thickness along the direction of the incident plane wave propagation and infinitely periodic along the other direction [Figure~\ref{fig:system}(b)]. The period along the $y$-axis is fixed at $a_y = 120$~mm. 
Unless it is stated otherwise, the thickness of the semi-infinite structure is $3$ pairs of the resonators since in Ref.~\cite{krasikova2023metahouse} it was demonstrated that such thickness is enough to experimentally observe pronounced stop-bands corresponding to the band-gaps of an equivalent infinite system [Figure~\ref{fig:system}(b)]. 
It might be argued that a structure with $3$ unit cells can not be called a truly periodic one and variation of parameters in the direction of wave propagation is not enough to claim it to be chirped. However, below it will be demonstrated that even properties of a structure with a single unit cell thickness can be linked to an infinitely periodic system.
For that three different configurations of the considered systems are investigated. The first type relies on the modification of unit cells rather than of the resonators. These are the structures in which position of the resonators are varied, such as the structures with the chirped distances and with super-cells (see Supplementary Information). The other types of the structures imply modification of the resonators via variation of slit widths or introduction of porous inserts for control of intrinsic losses. Though they might seem to be different, all of these configurations basically allow to modify the local coupling of the resonators extending the possibilities for bang-gap engineering and tuning of noise-insulating properties.

\subsection{Chirped distances}
\begin{figure}[htbp!]
    \centering
    \includegraphics[width=\linewidth]{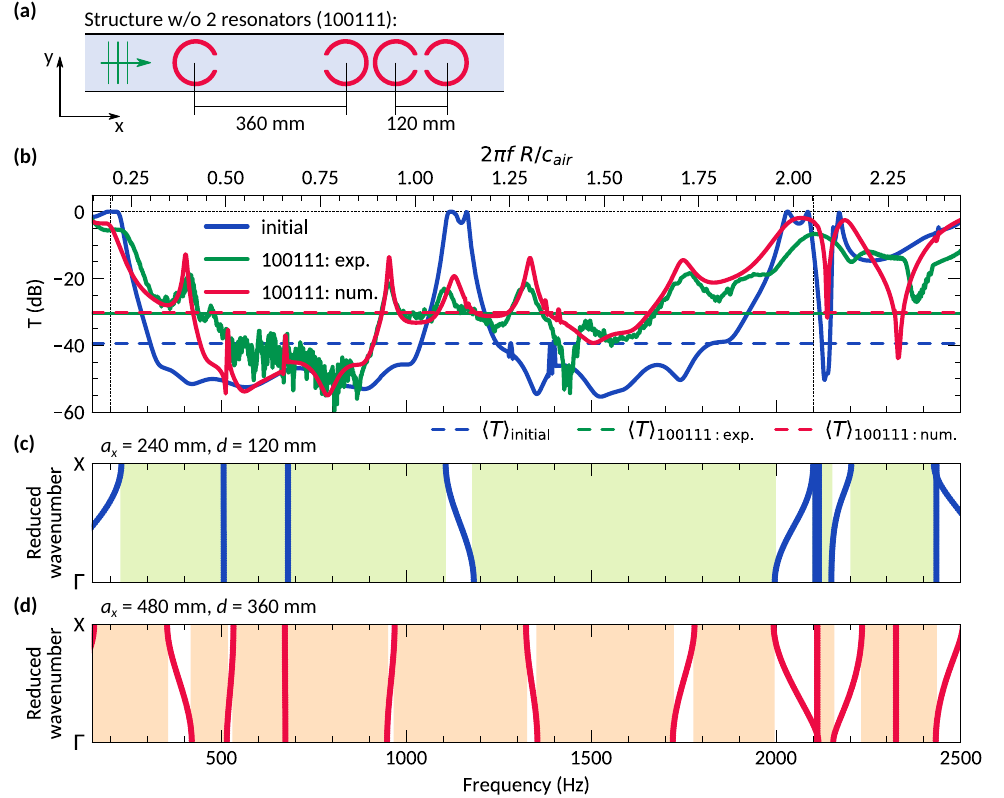}
    \caption{\textbf{Structure with the removed pair of resonators.} Schematic geometry of the considered structures for (a) optimized designs. The black lines along the direction of incident plane wave propagation are Floquet periodic boundaries. (b) Transmission spectra of the corresponding structures. Dashed horizontal lines correspond to the spectrally averaged transmission coefficient within the range $200$--$2100$~Hz.
    Band diagrams for the infinite structures equivalent to the structures with (c) $a_x = 240$~mm, $d = 120$~mm and (d) $a_x = 480$~mm, $d = 360$~mm. The shaded areas indicate the $\Gamma X$ band-gaps.}
    \label{fig:supercell_011000}
\end{figure}

One of the major factors affecting the noise-insulation properties is the local coupling between the resonators as the variation of other parameters do not result in the suppression of undesired resonances in the transmission spectra (see Supplementary Information). The coupling coefficient is mostly affected by the distance of the resonators and their slit widths. For simplicity, the investigation starts from the case when only one resonator is displaced and the number of pairs is reduced to $2$. More complicated structures are considered in the Supplementary Information.
To consider a particular example we choose the design with $d_1 = 360$~mm and $d_2 = 120$~mm [see \textbf{Figure~\ref{fig:supercell_011000}}(a)]. It could be noted that such a system can be formed via the removement of two resonators from the initial structure with $3$ pairs. This is an important remark since the selected structure represents one of the designs obtained with the help of the genetic algorithm optimization procedure (see Methods section). This is the reason why the structure is labeled as 100111 --- in this coding (1) stands for the resonator which was not removed and (0) stands for the removed resonator. In this notation the initial structure can be labeled simply as 111111.

As it is shown in Figure~\ref{fig:supercell_011000}(b), the stop-bands in both the calculated and measured transmission spectra in this case merge, resulting in a broad-band noise suppression. 
This enhancement can be explained via consideration of the equivalent infinite systems since the optimized structure is basically made of two pairs of coupled resonators with different spacing within a pair, namely $120$ and $360$~mm. Indeed, Figure~\ref{fig:supercell_011000}(c) and~\ref{fig:supercell_011000}(d) demonstrate that several eigenmodes of one structure lie within band-gaps of the other structure and vice versa. Therefore, the resonance near $1100$~Hz occurring in the spectrum of the initial structure is suppressed.
At the same time, other resonances occur in the spectra but they are not pronounced exactly due to the overlap between the eigenmodes and band-gaps of two structures. Nevertheless, the spectrally averaged transmission coefficient in this case is about $-30$~dB which is $10$~dB higher than for the initial structure as the dips in the spectra become less deep. 
Hence, the obtained results can be used as a basis to form a 
conclusion that overlap between the band-gaps and eigenmodes of structures with different unit cells may be a mechanism for enhancement of noise-insulating properties. These overlaps may be achieved via tuning of local coupling between the resonators.

\subsection{Chirped slit widths}
To develop the idea further, it should be noted that variation of the distances is not the only way to alternate the coupling between the resonators and it also can be done via the variation of slit widths. The last case might be more beneficial as the alternation of the slit widths results in a shift of individual Helmholtz resonances which provides an additional degree of freedom for tuning of the band structures. 

\begin{figure*}[htbp!]
    \centering
    \includegraphics[width=\linewidth]{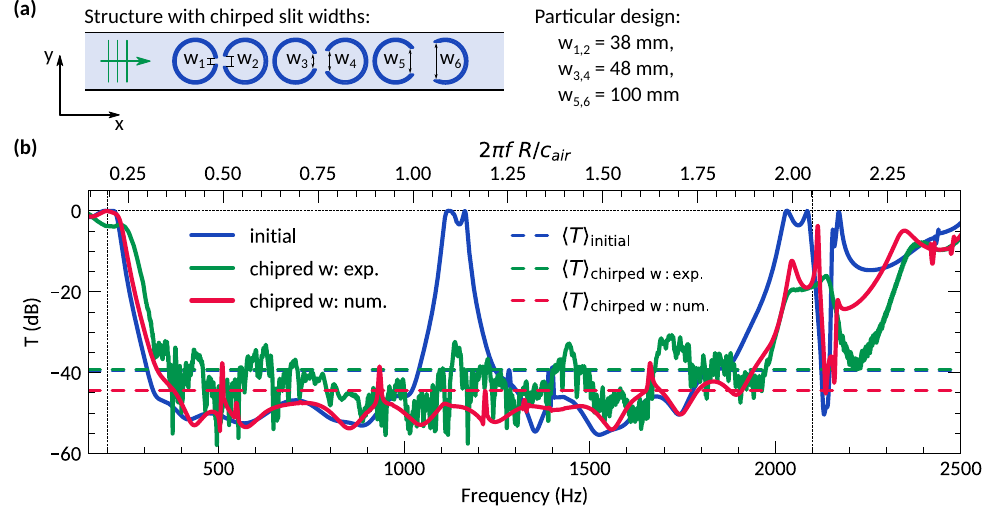}
    \caption{\textbf{Chirped slit widths.} (a) Schematic illustration of the considering system in which all resonators have different slit width $w$. (b) Transmission spectra for the initial case $w_{1-6} = 40$~mm and for the optimized structure with $w_{1,2} = 38$~mm, $w_{3,4} = 48$~mm and $w_{5,6} = 100$~mm. Dashed horizontal lines indicate spectrally averaged transmission coefficient within the range $200$ -- $2100$~Hz.}
    \label{fig:GA_w_chirp_spectra}
\end{figure*}

The schematics of the considered structure in which the slit width within each pairs are different is shown in \textbf{Figure~\ref{fig:GA_w_chirp_spectra}}(a). 
In general, all of the slit widths can be varied which makes the tuning of the system more difficult but at the same time does not provide any benefits as shown in the Supplementary Information. 
Figure~\ref{fig:GA_w_chirp_spectra}(b) demonstrates the transmission spectra for the structure with $w_{1,2} = 38$~mm, $w_{3,4} = 48$~mm and $w_{5,6} = 100$~mm. This design was obtained with the help of the genetic algorithm optimization procedure described in the Methods section. The stop-band in this case becomes nearly flat such that the experimentally obtained spectrally averaged transmission is about $-40$~dB with a deviation just $10.5$~dB. In addition to the ultra-broad flat stop-band, the structure with chirped slit width is thinner than the one in which distances between the resonators are tuned.

\subsection{Lossy resonators}
\begin{figure*}[htbp!]
    \centering
    \includegraphics[width=\linewidth]{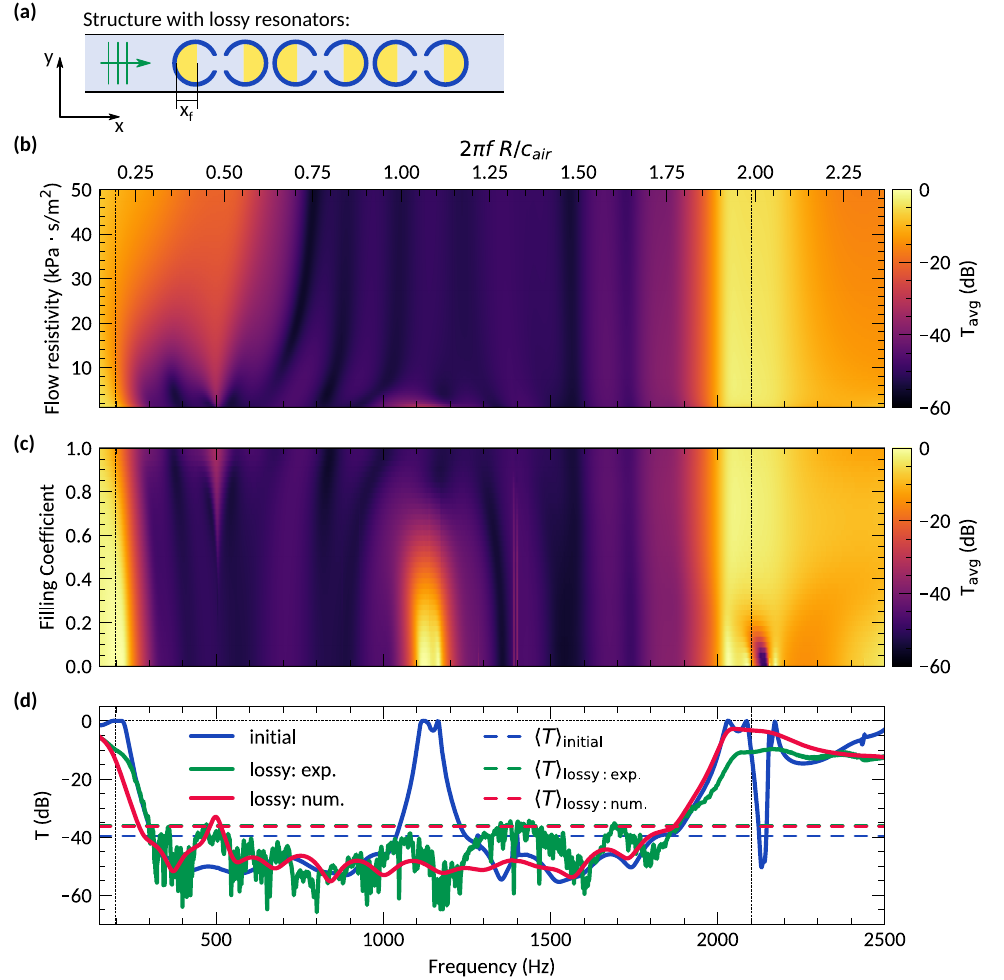}
    \caption{\textbf{Semi-infinite system with lossy resonators.} 
    (a) Schematic illustration of the system. The resonators are filled with poroacoustic material with the thickness $x_f$. (b) Transmission spectra as a function of the flow resistivity for the fully filled resonators. (c) Transmission spectra as a function of the filling coefficient $x_f/r$, where $r$ is the inner radius of the resonators. (d) Transmission spectra for the case when the inserts fully occupy the volume of the resonators. Dashed horizontal lines indicate spectrally averaged transmission coefficient within the range $200$--$2100$~Hz. The flow resistivity in panels (c) and (d) is $6$~kPa$\cdot$s/m$^2$.
    }
    \label{fig:poroacoustics}
\end{figure*}

For the above results variation of local coupling strength was performed via geometrical changes of the resonators or their positioning in the structures. However, no mechanisms of intrinsic losses were considered so far. In this section, we briefly discuss the case of lossy resonators formed via the addition of porous materials in their volume [see \textbf{Figure~\ref{fig:poroacoustics}}(a)]. On the one hand, porous materials themselves are widely used for noise-insulating systems, but on the other hand, they can significantly degrade the resonant properties of the resonators, which might deform the corresponding band-structures and lead to the narrowing or closure of the band-gaps. This statement is supported by Figure~\ref{fig:poroacoustics}(b), where transmission spectra of the periodic structures supplemented by materials with different flow resistivity are presented. The flow resistivity for the considered purposes is the only parameter defining the properties of porous materials as the Delany-Bazley-Miki model~\cite{miki1990acoustical,oliva2013sound} is used.

Indeed, additional losses might enhance the noise-insulating properties in case the relatively low values of flow resistivity (below $10$~kPa$\cdot$s/m$^2$), but the stop-band at $300$-$800$~Hz closes when the flow resistivity increases. This happens due to the suppression of the Helmholtz resonances caused by huge intrinsic losses [see Supplementary Information]. Since the wide band-gaps of the initial structure were the consequence of the strong coupling between the resonators the absence of the corresponding resonances prevents the formation of the associated band-gaps. Therefore, the result might be quite counter-intuitive as higher losses in this case provide better transmission since the properties of the locally resonant periodic structure are not utilized in full. At the same time, the absorption coefficient of the structure is low [see Supplementary Information] which also indicates that the noise-insulation properties are related to periodicity of the structure rather than losses in porous materials.

Therefore, for the subsequent considerations the flow resistivity is fixed at $6$~kPa$\cdot$s/m$^2$ reasonable for various foams~\cite{bies1980flow}. Then, the volume occupied by the porous insert is varied [see Figure~\ref{fig:poroacoustics}(a)]. The calculated transmission spectra [see Figure~\ref{fig:poroacoustics}(b)] indicate that the resonance near $1100$~Hz becomes suppressed when the filling coefficient reaches values of $0.6$ or more. This can be also seen in Figure~\ref{fig:poroacoustics}(c), where the numerically and experimentally obtained spectra for the case of the full filling are presented. The stop-band in this case is also nearly flat and actually wider than in the previous cases which is the merit of the porous inserts.
It should be noted that the differences between the calculated and measured spectra might result from the mismatch between the parameters of porous materials used for simulations and experiment, since the flow resistivity of the used foam rubber is unknown. 

\subsection{Discussion}
\begin{figure}[htbp!]
    \centering
    \includegraphics[width=\linewidth]{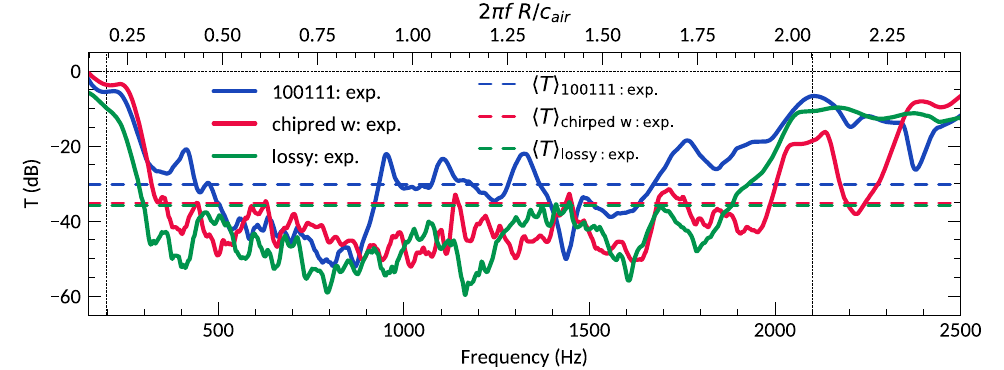}
    \caption{\textbf{Comparison of the measured transmission spectra for different designs.} The spectra are significantly smoothed for the better visualization. Dashed horizontal lines indicate spectrally averaged transmission coefficient within the range $200$--$2100$~Hz.}
    \label{fig:comparison_spectra}
\end{figure}

The presented results are summarized in \textbf{Figure~\ref{fig:comparison_spectra}} and Table~\ref{tab:comparison}, where the considered structures are compared. It might be mentioned that the spectrally averaged transmission coefficient is not sufficient to compare the structures as almost for all of them it takes similar values while the heights of the peaks and flatness of the stop-bands are different. Therefore, the standard deviation of the transmission coefficient is also considered as it allows us to evaluate the presence of undesirable resonances in the spectra.

Among the presented designs the structures with porous inserts and with the chirped slit width seem to be the most beneficial as they are characterized by the wide nearly flat stop-bands covering at least the range from $300$ to $2000$ where the transmission is below $-20$~dB. At the same time, the thickness of this structures remains the same as of the initial one. However, the structure with lossy resonators  obviously requires more materials. In addition, it might be argued that a porous materials of the same thickness as the considered structure could provide better noise-insulation. While this remark is reasonable, the benefit of the resonator-based structure is its ventilation. Despite small distances between the resonators the channels are not small enough to block air flow, though the flow distribution behind the structure is rather complicated (see Supplementary Information). Optimization of the ventilation performance lies beyond the scope of the work, but it is expected that a compromise between noise-insulation and ventilation properties could be found. 
In addition, the weight of the proposed structures might be significantly less than of the conventional ones, due to the less amount of materials required to achieve noise-insulating properties.

\begin{table*}[htbp!]
    \centering
    \caption{Comparison of the considered structures in terms of their transmission properties.}
    \label{tab:comparison}
    \newcolumntype{Y}{>{\centering\arraybackslash\hsize=.5\hsize}X}
    \newcolumntype{S}{>{\arraybackslash\hsize=.5\hsize}X}
    \newcolumntype{T}{>{\arraybackslash\hsize=.5\hsize}X}
    \begin{tabularx}{\linewidth}{TS||YY|YY}
        \multirow{2}{*}{Structure type} & & \multicolumn{2}{c|}{$200$--$2100$~Hz} & \multicolumn{2}{c}{$150$--$2500$~Hz} \\ 
        & & $\langle T \rangle$, dB & $T^{\mathrm{std}}$, dB & $\langle T \rangle$, dB & $T^{\mathrm{std}}$, dB \\
        \hline
        Initial & Ref.~\cite{krasikova2023metahouse} & -31.6 & 7.2 & -34.2 & 18.8 \\
        100111 & Figure~\ref{fig:supercell_011000} & -30.4 & 11.1 & -27.1 & 12.3\\
        Chirped w & Figure~\ref{fig:GA_w_chirp_spectra} & -39.3 & 10.5 & -35.2 & 13.6 \\
        Lossy & Figure~\ref{fig:poroacoustics} & -41.7 & 11.4 & -35.9 & 15.8 \\
    \end{tabularx}
\end{table*}

\section{Methods}
\subsection{Numerical calculations}
All numerical calculations are performed with in COMSOL Multiphysics via the Pressure Acoustics, Frequency Domain module. The schematic of the model is shown in Fig.~\ref{fig:methods}(a).
In order to avoid any reflections from the end of the waveguide perfectly matched layers were introduced. At the same time, Floquet periodic conditions were applied to the waveguide boundaries parallel to the incident wave propagation. 
It was also assumed that the boundaries of resonators are sound-hard walls which are justified by the fact that the impedance of plastics is several orders higher than that of air~\cite{selfridge1985approximate}.
For the case of the resonators, the porous material was introduced via the Poroacoustics feature using the Delany-Bazley-Miki model and Delany-Bazley constants.
The incident wave was introduced as the Background Pressure Field with amplitude $1$~Pa.
Then, the transmission spectra were calculated as 
\begin{equation}
    T = 20 \log_{10} \left( \frac{p_{\mathrm{tr}}}{p_{\mathrm{ref}}} \right),
    \label{eq:transmission}
\end{equation}
where $p_{\mathrm{tr}}$ is a root mean square of pressure amplitudes evaluated at several arbitrary sampling points located behind the structure and $p_{\mathrm{ref}}$ is the corresponding root mean square of the incident pressure amplitude.
Despite the fact that the experimental results are reliable only within the range from $200$ to $2100$~Hz, as explained below, numerical computations are provided for a bit wider range $150$ -- $2500$~Hz in order to better understand the results of the optimization procedures. All graphs with transmission spectra are also substituted by the normalized frequency scale as the considered systems are scalable.
For better visualization of the spectra, all of them are slightly smoothed via the Savitzky-Golay filter of order $2$ and with widow width $11$. It is also should be noted that thermo-viscous losses are not accounted for during the calculations which might be a reason for some discrepancies between the numerical and experimental results.

\subsection{Experimental measurements}
\begin{figure}[htbp!]
    \centering
    \includegraphics{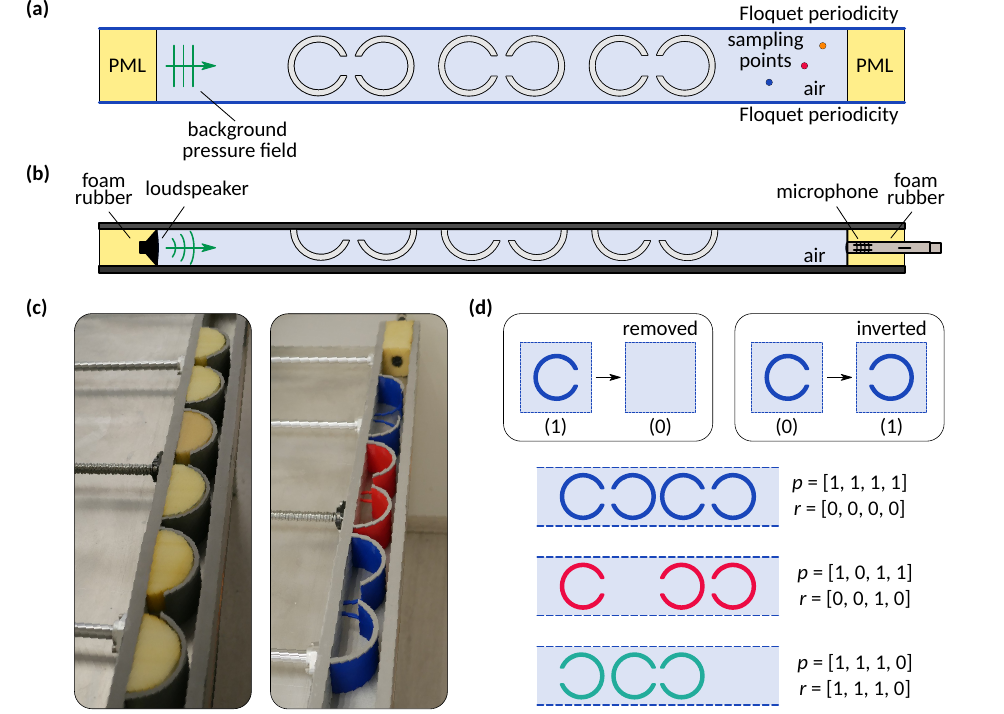}
    \caption{\textbf{The utilized methods.} (a) Schematic illustration of the considered numerical model consisting of a waveguide with periodic boundary conditions and PML layers at the ends. Pressure is calculated at three arbitrary sampling points behind the structure and then evaluated as a root mean square of these values. 
    (b) Schematics of the experimental setup representing a waveguide with square cross-section. The loudspeaker and the microphone are embededd in the porous material inserted at the ends of the waveguide. Instead of the full resonators, only the halves of them are used which is equivalent due to the symmetry of the structure. (c) Photo of the samples with chirped slit width and with porous inserts. (d) Schematics of the operations used in the binary genetic algorithm for optimization of the structures. Two operations are performed: removal of a resonator and its inversion. Then the structures can be described by the binary matrices indicating the performed operations.}
    \label{fig:methods}
\end{figure}

All measurements were conducted inside a waveguide with square cross-section as shown schematically in Fig.~\ref{fig:methods}(b). The cross-section side is equal to $60$~mm and the full length of the waveguide is $1350$~mm. All boundaries are considered to be hard sound boundaries since the walls and the bottom of the waveguide are made of $15$~mm thick aluminium and the lid is made of $6$~mm thick plexiglass. The interior of the waveguide is insulated by foam rubber pieces inserted at both ends of the waveguide. The thickness of each piece is $150$~mm. 
The acoustic waves were generated by the loudspeaker (Visaton BF 45/4 with Yamaha YDA138-E based amplifier) located at one end of the waveguide while the recording of the signal was done with the microphone (Rode NTG-4) placed at the other end.
Both the speaker and the microphone were embedded into the foam rubber inserts.
Generation and recording of signals were controlled via the USB audio interface Roland Rubix22 connected to a personal computer and hand-made software based on the Python programming language and the sound device module~\cite{sounddevice}.

The sound hard boundary condition allowed us to imitate periodicity along the directions perpendicular to the waveguide axis (direction of the incident wave propagation). The imperfectness of this assumption may be the main reason for the small discrepancies between the numerical and the experimental results.
The operational spectral range $200$ -- $2100$~Hz was selected in such a way that the cross-section side is smaller than half of the wavelength, which allows to preserve the symmetry of the unit cell and imitates infinite periodicity.
The symmetry of the considered structures also allowed using halved resonators instead of the full ones, which helped to adjust the system to the considered spectral range.

The samples were manufactured by 3D printing (PLA filament) [see Fig.~\ref{fig:methods}(c)]. The resonators were provided with small place-holders to simplify their positioning inside the waveguide. The porous inserts were made of foam rubber with the flow resistivity assumed to be of the order $6$~kPa$\cdot$s/m$^2$, but this value was not verified.

For each measurement, the chirped signal was generated several times and the obtained transmission spectra were averaged in Fourier space. The sampling frequency was set to $44100$~Hz. The transmission coefficient is defined in a similar way as for numerical calculations [see eq.~\eqref{eq:transmission}] but with the correction that $p_{\mathrm{ref}}$ is the reference amplitude (without structure) and $p_{\mathrm{tr}}$ is the amplitude of the transmitted signal.
The obtained spectra are smoothed in the same way as the numerically obtained ones.

\subsection{Genetic algorithm optimization}
The optimization procedure was developed on the basis of classical genetic algorithms~\cite{mitchell1998introduction}.
The aim of the procedure was to minimize the cost function defined as the linear combination of the spectrally averaged transmission coefficient $\langle T\rangle$, its standard deviation $T^{\mathrm{std}}$ and the maximal value $T^{\mathrm{max}}$ within a specified spectral range:
\begin{equation}
    \mathcal{C} = a_1 \langle T\rangle + a_2 T^{\mathrm{std}} + a_3 T^{\mathrm{max}}
\end{equation}
where $a_{1-3}$ are weights adjusted during the optimization procedure. 
In this case, it is possible not only to achieve low averaged transmission but also to avoid undesired resonances and achieve flat stop-bands. An extensive discussion of the objective functions and their suitability can be found, for instance, in Ref.~\cite{marburg2002developments}.
The algorithm parameters were selected arbitrarily. For each case, the population size was fixed at $20$ and the selection procedure was performed via the roulette wheel method with the selection pressure $0.1$. 
Implementation of the crossover implied random usage of the single point, double point or uniform crossover operators, with the probabilities of choice $0.2$, $0.2$ and $0.6$, respectively.
The mutation rate was set to $0.1$ and for the real coded algorithm, the normal distribution with standard deviation $0.1$ was used. 
For the case of the slit width variation, the values were bounded by $1$ and $26.5$~mm (or $R/2$). The distances between the resonators within a pair were limited by $106$ and $720$~mm. As for the structures with supercells, the binary genetic algorithm was used in this case. Each resonator was labelled by ($x$, $y$), such that $x$ defines the presence of a resonator which is either present ($1$) or removed ($0$). Then, $y$ defines whether the resonator is rotated by $180$ degrees ($1$) or not ($0$). Therefore, each structure can be described by a binary matrix of size $2  \times \text{number of resonators}$, such that one half of the matrix defines the presence of resonators and another half -- their inversion [see Fig.~\ref{fig:methods}(d)].

\section{Conclusions}
We demonstrated that the noise-insulating properties of periodic structures based on Helmholtz resonators can be tuned via variation of local coupling between the resonators. In particular, such tuning can be done via chirping of parameters, such as the distance between the resonators or their slit width. Also, the local coupling can be altered via introduction of porous inserts into the bodies of the resonators allowing control of the intrinsic losses. As a result, the presented tuning methods allow merging of the stop-bands associated with the overlap of band-gaps and eigenmodes of the equivalent infinite structures. This leads to wide stop-bands, with low average transmission, and without any strong resonant transmission peaks.

\medskip
\textbf{Supporting Information} \par 
Supporting Information is available from the Wiley Online Library or from the author.

\medskip
\textbf{Acknowledgements} \par 
The authors thank Yuri Shchelokov and Aleksandr Kalganov for the help with assembly and alignment of the experimental setup. The authors also thank Alexey Dmitriev for fruitful discussions. 
M.Kr. acknowledges the German Academic Exchange Service DAAD, Research Grants - Bi-nationally Supervised Doctoral Degrees, A.L. acknowledges the Ministry of Science and Higher Education of the Russian Federation (Project No. 075-15-2022-1120), A.B. and S.K. acknowledge the support from Priority  2030  Federal  Academic  Leadership  Program.

\medskip
\textbf{Author contributions} \par 
M.Kr., S.K. and A.P. provided the numerical calculations. A.P., M.Kr., M.K. and A.L. performed the experimental measurements. S.K. developed the optimization algorithms. M.Kr., S.K., Y.B. and A.B. proposed and developed the idea of the project. A.M., Y.B., D.P., S.M. and A.B. provided guidance on all aspects of the work. S.M. and A.B. supervised the project. All authors contributed to writing and editing of the manuscript.

\medskip
%
\bibliographystyle{MSP}
\bibliography{bibliography}

\clearpage
\appendix
\begin{center}
    {\huge Supplementary Information}
\end{center}

\section{Chirped distances between the pairs}
\begin{figure}[htbp!]
    \centering
    \includegraphics[width=\linewidth]{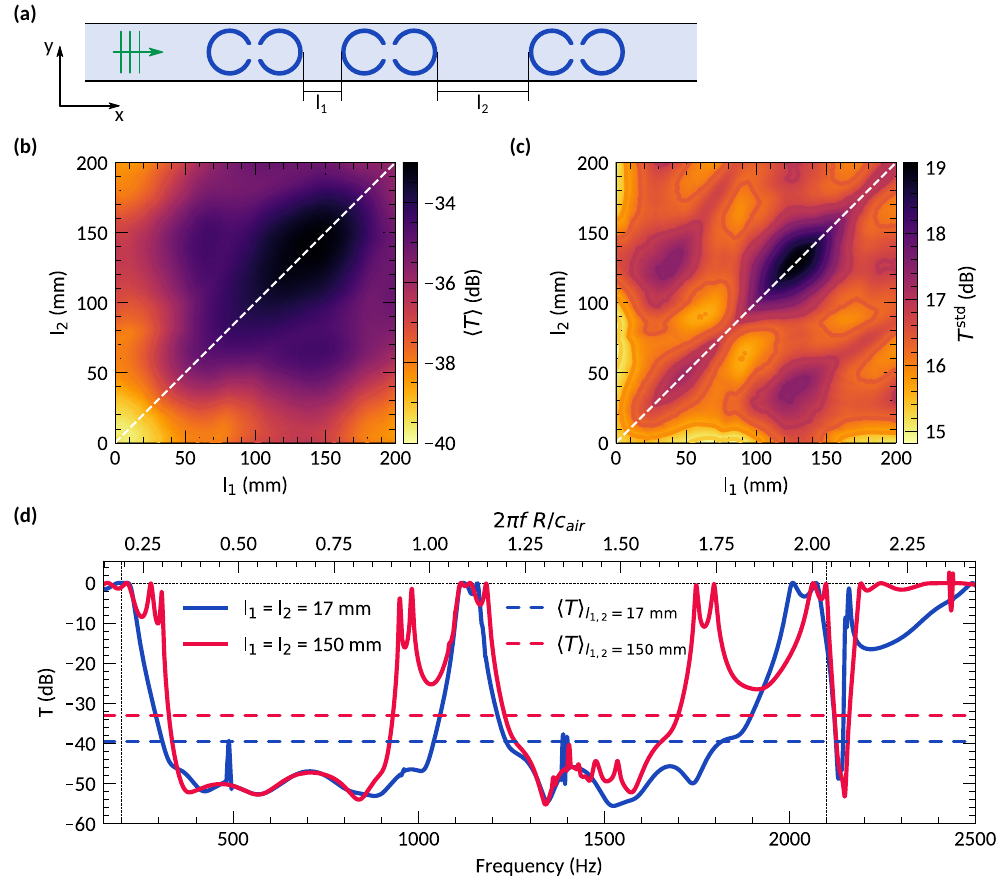}
    \caption{\textbf{Variation of distances between the pairs of resonators.} (a) Schematic illustration of the considered structure. (b) The spectrally averaged transmission coefficient and (c) its standard deviation as functions of distances between the pairs of the resonators $l_{1,2}$ within the range $200$~--~$2100$~Hz. White dashed lines correspond to the $l_1 = l_2$ line. (d) Transmission spectra for the structures with $l_1 = l_2 = 17$~mm and $l_1 = l_2 = 150$~mm.}
    \label{fig:chirped_3pairs_l1_l2}
\end{figure}

As it was discussed in the main text (see Introduction section), chirping of a structure may lead to the broadening of band-gaps.
Therefore, the investigations start from the simplest type of chirped structures in which the distances between the pairs of the resonators $l_{1,2}$ are varied [see Fig.~\ref{fig:chirped_3pairs_l1_l2}(a)], while distances between the centers of the resonators within each pair are fixed at the particular value $d = 120$~mm. 
Due to the reciprocity, it is expected that the interchange of $l_1$ and $l_2$ values will not affect the scattering properties of the structure. Hence, the map of the spectrally averaged transmission coefficient shown in Fig.~\ref{fig:chirped_3pairs_l1_l2}(b) as well as the corresponding standard deviation in Fig.~\ref{fig:chirped_3pairs_l1_l2}(c) are symmetric with respect to the $l_1 = l_2$ line.

\begin{figure}[htbp!]
    \centering
    \includegraphics[width=\linewidth]{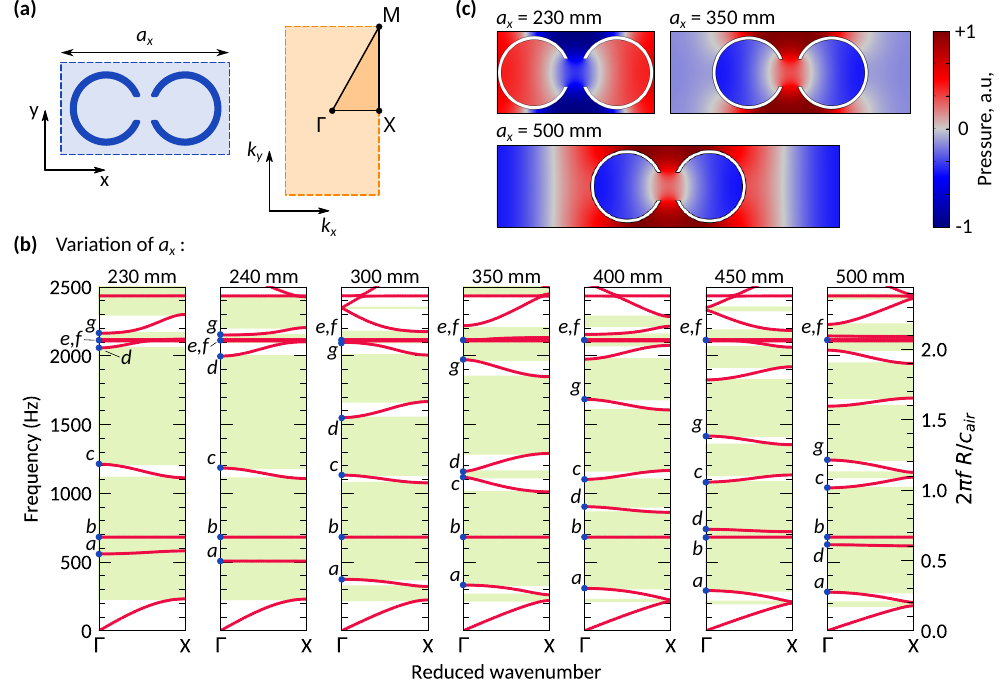}
    \caption{\textbf{Band structure of the system with varied width of the unit cell}. (a) Geometry of the unit cell. (b) Evolution of a band structure with the increase of the unit cell width. Green shaded areas correspond to $\Gamma X$ band-gaps. Letters $a$--$g$ are used for indication of the eigenmodes. (c) Field distributions corresponding to the eigenmode $c$ at the $\Gamma$-point.}
    \label{fig:band_structure_ax}
\end{figure}

The variation of the averaged transmission coefficient in this case is rather insignificant and the difference between the largest and the smallest values is less than $7$~dB. It is also can be mentioned that all transmission spectra are characterized by a resonance near $1100$~Hz similar to the examples shown in Fig.~\ref{fig:chirped_3pairs_l1_l2}(c).
This result can be explained via consideration of the corresponding infinite systems consisting of the unit cells with different widths $a_x$ [see Fig.~\ref{fig:band_structure_ax}(a)]. Figure~\ref{fig:band_structure_ax}(b) demonstrates the evolution of the band structure with the change of $a_x$. As it might be expected, an increase in the unit cell width results in a shift of eigenmodes toward the lower frequencies.

\begin{figure}[htbp!]
    \centering
    \includegraphics[width=\linewidth]{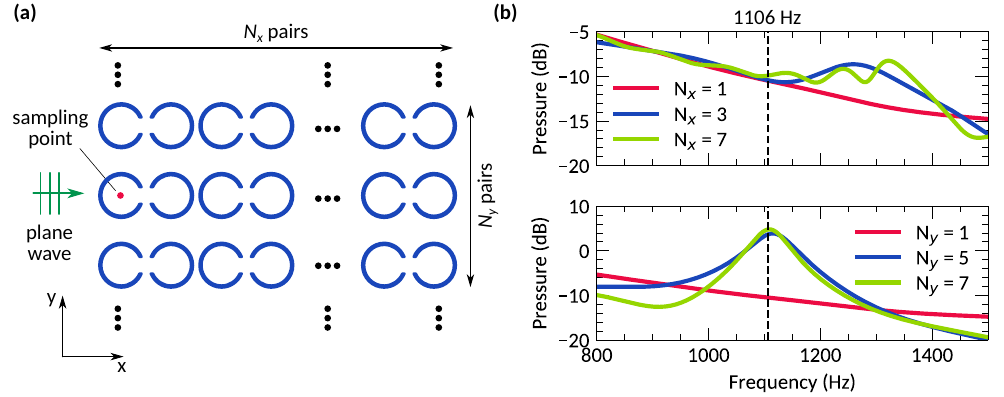}
    \caption{\textbf{Excitation of resonances in a finite-size system.} (a) Schematic illustration of a finite-size system consisting of $N_x$ pairs along the direction of the incident wave propagation and $N_y$ pairs along the other direction. Transmission spectra of the structures with (b) $N_y = 1$ and varied $N_x$ and (c) $N_x = 1$ with varied $N_y$.}
    \label{fig:mode_c_excitation}
\end{figure}

Nevertheless, at least two modes are almost unaffected and remain at nearly the same position. One of these eigenmodes labeled as $b$ occurs as a flat band corresponding to an anti-symmetric resonance which is not coupled with the other modes and hence the field is localized inside the resonators.
Another mode which is located near $1100$~Hz and labeled as $c$ might be considered as  a structural mode hybridized with the modes of the resonators as the field distributions shown in Fig.~\ref{fig:band_structure_ax}(c) suggest.
The nature of this eigenmode also might be checked via calculations of pressure inside one of the resonators in a finite-size system with the varied number of pairs. If this mode is a feature of the resonators themselves, the resonance should be excited even for the case of a single pair. However, in Fig.~\ref{fig:mode_c_excitation} it is shown that the resonance is excited only when additional pairs are introduced. 

\begin{figure}[htbp!]
    \centering
    \includegraphics[width=0.95\linewidth]{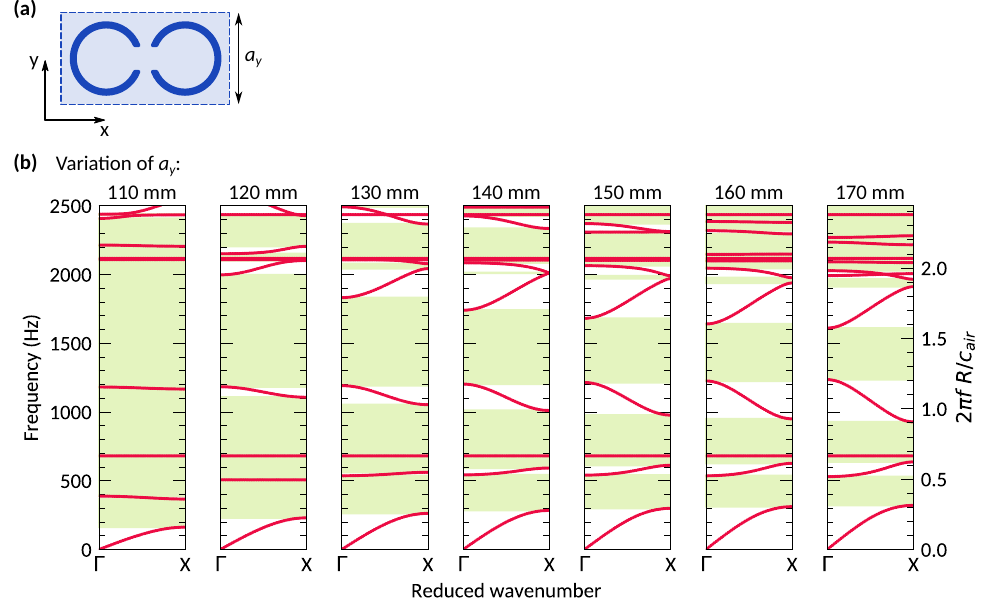}
    \caption{\textbf{Evolution of the band-structure with the change of the unit cell height.} (a) Geometry of the unit cell. (b) Evolution of a band structure with the increase of the unit cell height. Green shaded areas correspond to $\Gamma X$ band-gaps.}
    \label{fig:band_structure_ay}
\end{figure}

\begin{figure}[htbp!]
    \centering
    \includegraphics[width=0.95\linewidth]{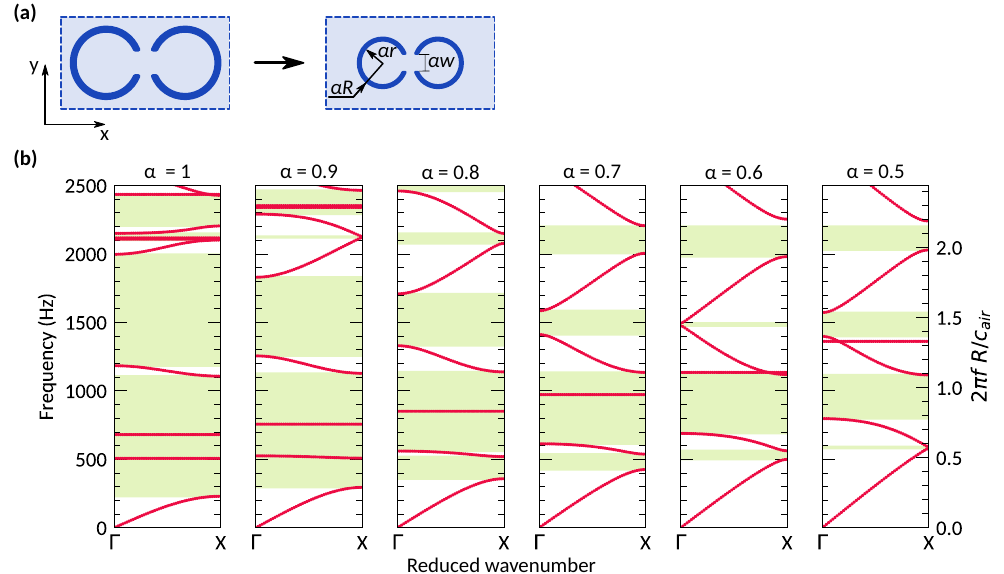}
    \caption{\textbf{Evolution of the band-structure with the scaling of the resonators.} (a) Geometry of the unit cell. (b) Evolution of a band structure with the scaling of the resontors. Green shaded areas correspond to $\Gamma X$ band-gaps.}
    \label{fig:band_structure_scale}
\end{figure}

It should be noted that position of this eigenmode is rather stable against the variation of the unit cell parameters. For instance, increase of the unit cell height [see Fig.~\ref{fig:band_structure_ay}] or scaling of resonators [see Fig.~\ref{fig:band_structure_scale}] results only in the increase of its incline, but not the spectral shift. Hence, the associated resonance in the transmission spectra can be observed for all considered combinations of $l_1$ and $l_2$. In order to change this situation, the position of this eigenmode has to be shifted which can be done via changes of the local coupling strength between the resonators.


\section{Chirped distances within the pairs}
As one of the parameters affectin the local coupling is the distance within the pair of the resonators, it is reasonable to start from the tuning of this parameter. 
To begin with a most simple case, the number of the pairs was reduced to $2$, such that the distance between the resonators of the first pair $d_1$ is varied while the distance within the second pair is fixed at $120$~mm, as previously. 

\begin{figure}[htbp!]
    \centering
    \includegraphics[width=0.98\linewidth]{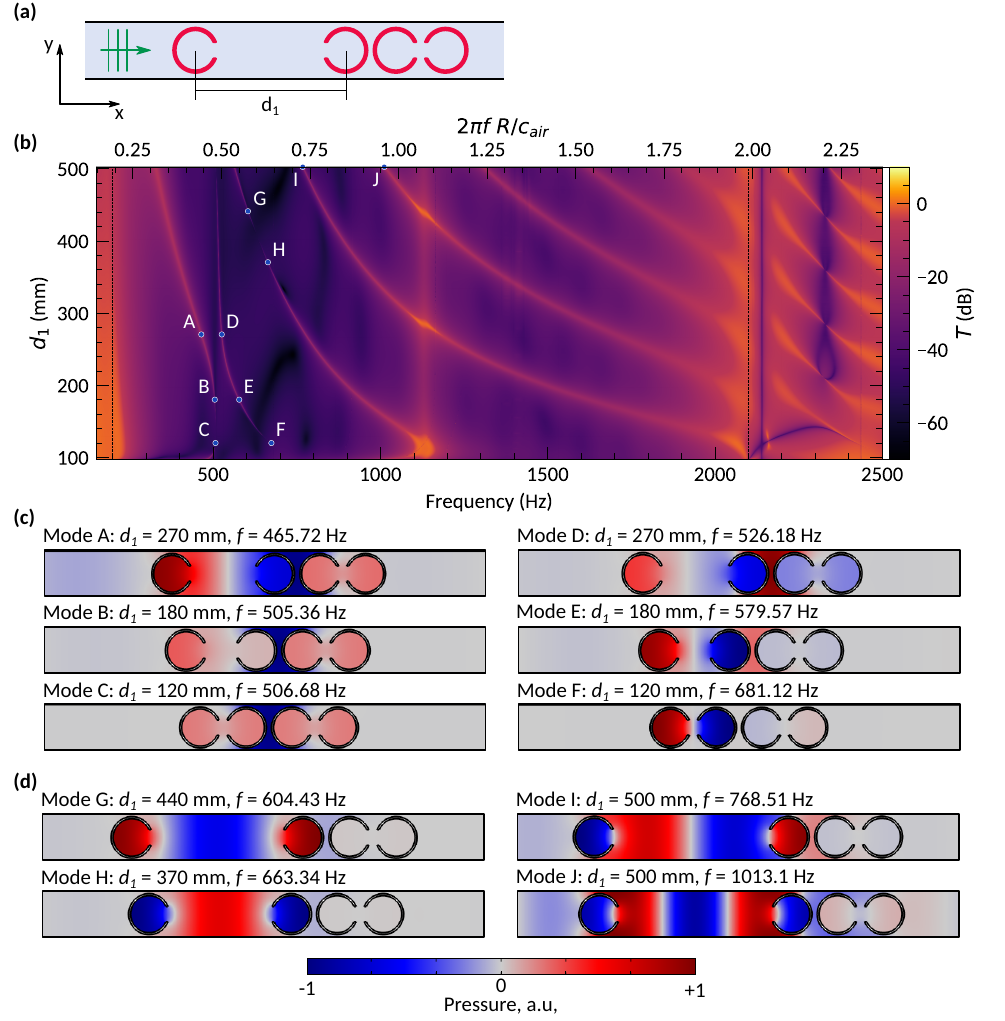}
    \caption{\textbf{Transmission of the structure with two pairs of resonators.} (a) Schematic illustration of the considered structure consisting of two pairs of the strongly coupled resonators. (b) Transmission coefficient as a function of frequency and distance between the resonators of the first pair $d_1$. (c),(d) Field distributions corresponding to modes labelled as $A$--$J$ on the transmission spectra map.}
    \label{fig:supercell_011000_map}
\end{figure}

Again, the resonance near $1100$~Hz is always present however in this case it is not pronounced except the regions of overlapping with other modes. The fact that such an intersection occurs at $d_1 = 120$~mm is the reason why changes of $l_1$ and $l_2$ do not result in the deformation of the resonance.
Another important remark should be made about the region in-between $400$ and $700$~Hz. The modes in this region correspond to the in-phase and out-of-phase excitation of Helmholtz resonators.
The mode labeled as $A$-$B$-$C$ in Fig.~\ref{fig:supercell_011000_map}(b) corresponds to the case when resonators of the second pair are excited in-phase, while the excitation of the first pair changes from the out-of-phase (point $A$) to the in-phase (point $C$) with the decrease of $d_1$.
In a similar manner, the mode $D$-$E$-$F$ is characterized by an out-of-phase excitation of the first pair and the change from the in-phase to out-of-phase excitation of the second pair. The corresponding field distributions are shown in Fig.~\ref{fig:supercell_011000_map}(c). 
Other modes occurring at higher frequencies might be considered to be of the Fabri-Perout type as the distance between the resonators becomes larger than the wavelength [see Fig.~\ref{fig:supercell_011000_map}(d)].

\begin{figure}[htbp!]
    \centering
    \includegraphics[width=\linewidth]{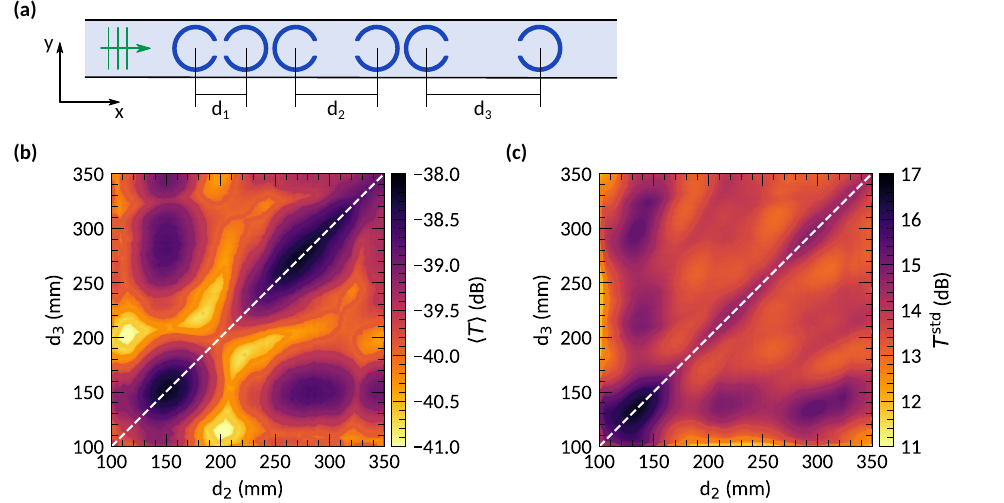}
    \caption{\textbf{Variation of distances within two pairs of the resonators.} (a) Schematic illustration of the considered system in which distances $d_2$ and $d_3$ are varied. (b) The spectrally averaged transmission coefficient and (c) its standard deviation as functions of distances $d_2$ and $d_3$ within the range $200$~--~$2100$~Hz. White dashed lines correspond to the $d_2 = d_3$ line.}
    \label{fig:chirped_3pairs_d2_d3_maps}
\end{figure}

In any case, the transmission spectra for $d_1 = 360$~mm (see Main Text) are characterized by the suppressed resonances near $1100$~Hz, which is a result of the overlap between bend-gaps and eigenmodes of the associated infinite structures.
This statement can be tested for a more complicated case, when $3$ pairs of the resonators are considered and the distances within two of them are varied [see Fig.~\ref{fig:chirped_3pairs_d2_d3_maps}(a)]. Similarly to the case of the varied $l_{1,2}$ the maps of spectrally averaged transmission coefficient and its standard deviation [see Figs.~\ref{fig:chirped_3pairs_d2_d3_maps}(b) and~\ref{fig:chirped_3pairs_d2_d3_maps}(c)] are symmetric with respect to $d_2 = d_3$ line due to reciprocity of the system. On the first glance, the results are not quite impressive as the averaged transmission remains almost the same for all values of $d_2$ and $d_3$. At the same time, the standard deviation is typically below $15$~dB suggesting that the resonances in the spectra might be more pronounced. It is also should be noted that the minima of the spectrally averaged transmission and its deviation occur aside from the line $d_2 = d_3$. This is an indication of the fact that chirped geometry is more beneficial for noise suppression than the initial one.

\begin{figure}[htbp!]
    \centering
    \includegraphics[width=\linewidth]{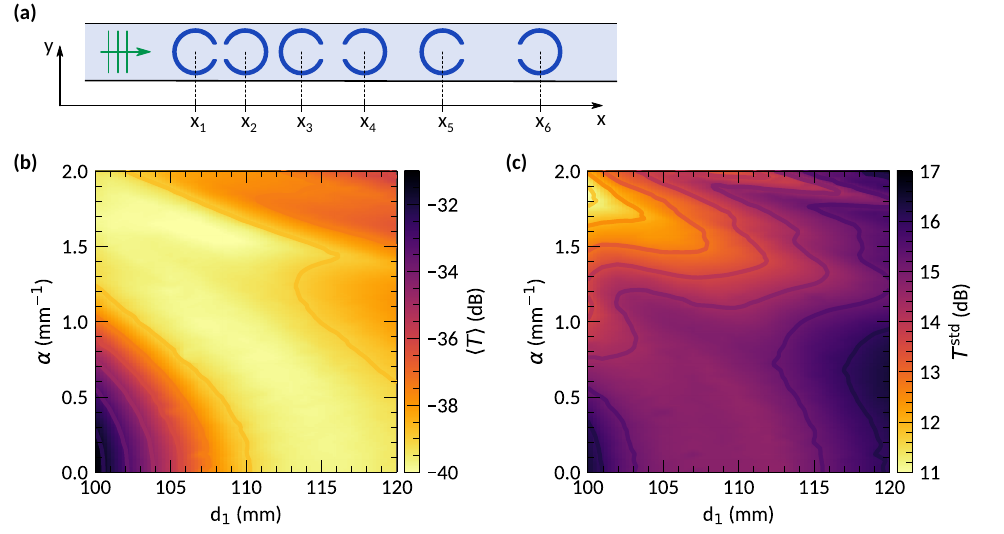}
    \caption{\textbf{Structure with the chirped distances between the resonators.} (a) Schematic illustration of the considered structure in which coordinates of the resonators $x_i$ follow the gauusian function $x_i = x_{i-1}+d_1 \exp(x_{i-1}^2 \alpha^2)$, such that $i$ is the number of a resonator and $\alpha$ is the chirp parameter. (b) The spectrally averaged transmission coefficient and (c) its standard deviation as functions of distances $d_1$ and $\alpha$ within the range $200$~--~$2100$~Hz.}
    \label{fig:chirped_3pairs_d1_alpha_map}
\end{figure}

\newpage
Before going into consideration of the transmission spectra the system is modified even further, such that all distances are varied according to the Gaussian function
\begin{equation}
    x_i = x_{i-1}+d_1 \exp(x_{i-1}^2 \alpha^2), 
\end{equation}
where $x$ is the coordinate of the $i$-th resonator, $d_1$ is the distance between the resonators within the first pair and $\alpha$ is a chirp parameter [see Fig.~\ref{fig:chirped_3pairs_d1_alpha_map}(a)]. In this case the spectrally transmission coefficient varies between $-32$ and $-40$~dB for the considered values of $d_1$ and $\alpha$ [see Fig.~\ref{fig:chirped_3pairs_d1_alpha_map}(b)]. The standard deviation of the transmission coefficient might be below $15$~dB for rather large region of $d_1$ and $\alpha$ values [see Fig.~\ref{fig:chirped_3pairs_d1_alpha_map}(c)].

\begin{figure}[hbp!]
    \centering
    \includegraphics[width=\linewidth]{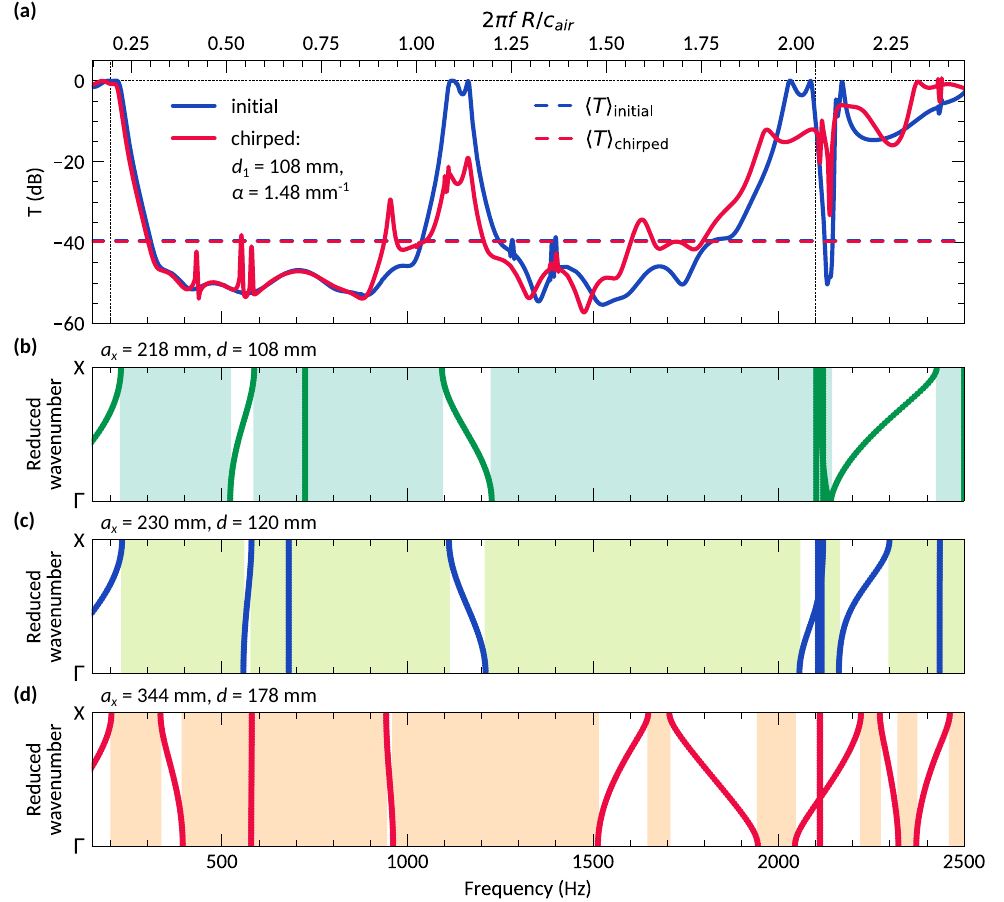}
    \caption{\textbf{Transmission of the structure with the chirped distances between the resonators.} (a) Transmission spectra of the chirped structure with $d_1 = 108$~mm and $\alpha = 1.48$~mm$^{-1}$ in comparison with the spectrum for the initial structure. Dashed horizontal lines correspond to the spectrally averaged transmission coefficient within the range $200$--$2100$ Hz. Band structures corresponding to the infinite systems with the unit cells characterized by (b) $a_x = 218$~mm, $d = 108$~mm, (c) $a_x = 230$~mm, $d = 120$~mm and (d) $a_x = 344$~mm, $d = 178$~mm. Shaded areas indicate the corresponding $\Gamma X$ band-gaps.}
    \label{fig:chirped_3pairs_d1_alpha_spectra}
\end{figure}

\newpage
To consider a specific example, we select the design with $d_1 = 108$~mm and $\alpha = 1.48$~mm$^{-1}$ characterized by the spectrally averaged transmission coefficient $-40$~dB with the deviation about $12.5$~dB. The corresponding spectrum shown in Fig.~\ref{fig:chirped_3pairs_d1_alpha_spectra}(a) is quite similar to the initial one but with a suppressed resonance near $1100$~Hz. Again, this can be explained via consideration of the equivalent infinite systems, since the chirped structure can be considered as the one made of three different unit cells. In this particular example, the width of the unit cells and the distance between the resonators are $a_x = 218$~mm, $d = 108$~mm for the first unit cell, $a_x = 230$~mm and $d = 120$~mm for the second one and $a_x = 344$~mm, $d =178$~mm for the third one. The band structures of the corresponding systems are shown in Figs.~\ref{fig:chirped_3pairs_d1_alpha_spectra}(b)--~\ref{fig:chirped_3pairs_d1_alpha_spectra}(d). Clearly, almost all eigenmodes of one structure lie in the range of band-gaps of two other structures. For instance, the mentioned mode at $1100$~Hz is present for unit cells with $a_x = 218$~mm and $a_x = 230$~mm, but for the unit cell with $a_x = 344$~mm it shifts towards $950$~Hz so at $1100$~Hz there is a band-gap. Therefore, in the transmission spectra the resonance at $1100$~Hz is not pronounced and additional resonance near $950$~Hz occurs. The resulting stop-band in this case can not be characterized as a flat one, but nevertheless, it covers the range $250$ -- $2150$~Hz in which the transmission is below $-10$~dB.

This result confirms the initial statement that chirping of the local coupling between the resonators may result in the broadening of stop-bands of transmission spectra which might be associated with the overlap of band-gaps and eigenmodes of the equivalent structures.

\section{Supercells}
\label{sec:supercells}

In general, the presented approach implies a combination of different unit cells in a manner allowing the overlap of band-gaps with eigenmodes. 
The presented results so far imply modification of geometric parameters without changes of the unit cell type, which also might be a beneficial strategy. For that elements in the unit cells can be changed. However, in order to not move too far away from the initial structure, we form new unit cells using simple operations of removal of a resonator or its inversion (rotation by $180$ degrees). 
The arbitrary angle of rotation is not considered solely for the sake of simplicity and reduction of the number of calculations. 
\begin{figure}[htbp!]
    \centering
    \includegraphics[width=0.9\linewidth]{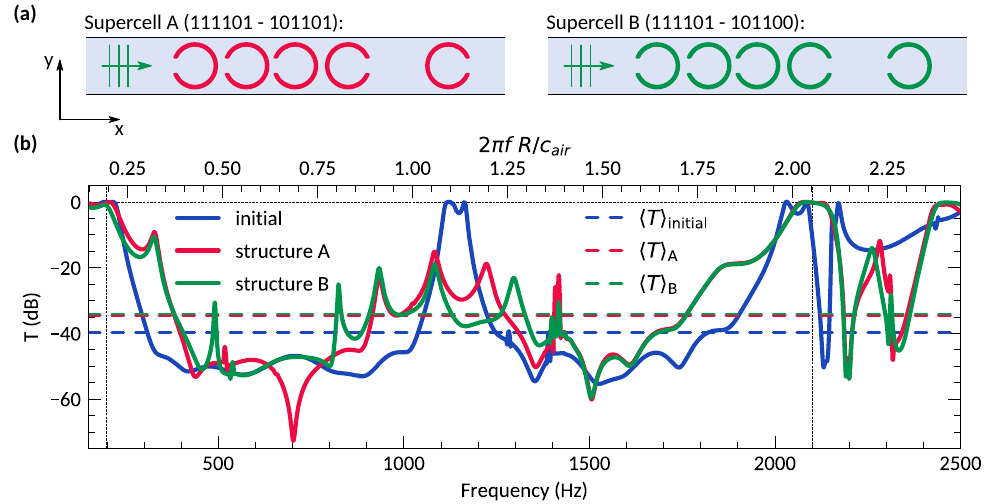}
    \caption{\textbf{Structures with super-cells.} (a) Schematic illustration of the considered super-cells. The labels given in brackets is separated by on two parts. The first part indicates presence ($1$) or absence ($0$) of a resonator and the second one -- whether it is inverted ($1$) or not ($0$). (b) Corresponding transmission spectra compared with the initial structure (111111 - 000000). Dashed horizontal lines indicate spectrally averaged transmission coefficient within the range $200$ -- $2100$~Hz.}
    \label{fig:GA_supercell_spectra}
\end{figure}
In this case, a row of $3$ pairs of the resonators can be considered as a single super-cell, which is modified by the mentioned operations.

Basically, the mentioned operations allow a change in the coupling coefficient via the transformation of a unit cell with a pair of resonators to a unit cell with a single resonator or via change of the distance between the resonators. Obviously, much more complicated designs of super-cells are possible, but they are not considered in this work, though might be even more beneficial.
Figure~\ref{fig:GA_supercell_spectra}(a) demonstrates two of the most beneficial designs obtained with the help of genetic algorithms (see Methods section of the main text). One of them consists of the unit cells with single resonators with necks turned toward the direction of the incident wave propagation and two resonators, with a gap between them, looking to the other direction. Another structure is nearly the same except the fact that the last two resonators face each other. The transmission spectra of these structures also resemble each other with the difference in the position of some resonances. However, in both cases the resonances within the range $200$ -- $2100$~Hz are not pronounced, again due to the overlap of bang-gaps and eigenmodes of the structures with different unit cells.
The spectrally averaged transmission in this case is about $-35$~dB, which is higher than for the initial structure but with the benefit of the suppressed resonances. In addition, the stop-band at higher frequencies, namely $2200$ -- $2400$~Hz, is also wider and more pronounced than for the initial structure. However, this design strategy seems to be not the optimal ones as other structures demonstrate better performance due to the higher tunability.

\section{Chirped slit widths}
\begin{figure}[hbp!]
    \centering
    \includegraphics[width=\linewidth]{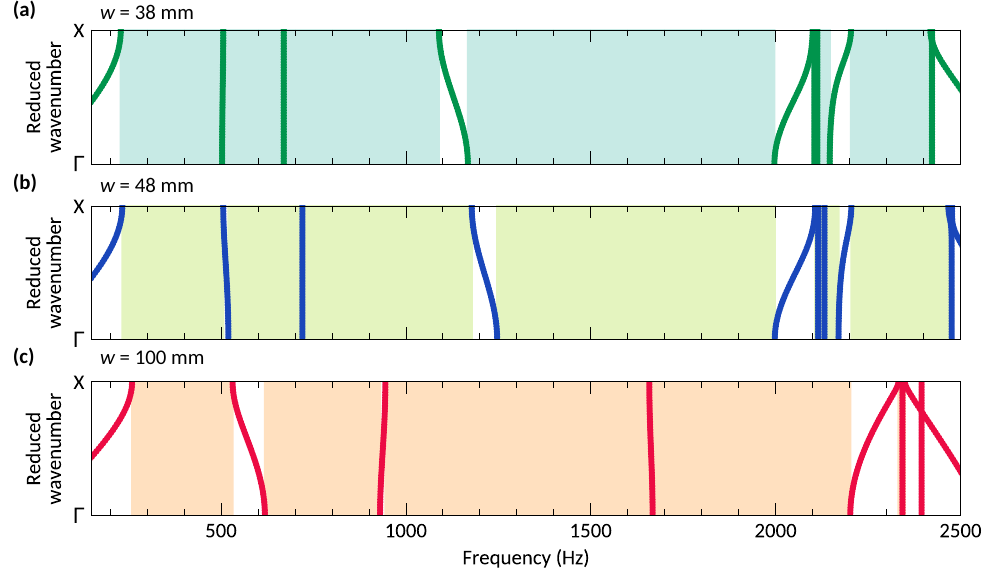}
    \caption{\textbf{Eigenmodes of structures with chirped slit widths.} Band structures corresponding to the infinite systems with the unit cells characterized by (a) $w = 38$~mm, (b) $w = 48$~mm, and (c) $w = 100$~mm. Shaded areas indicate the corresponding $\Gamma X$ band-gaps.}
    \label{fig:GA_w_pair_band}
\end{figure}

In the main text it is stated that variation of slit widths also can lead to the enhancement of the noise-insulating properties due to the merging of the stop-bands.
As Fig.~\ref{fig:GA_w_pair_band} shows, this is also a consequence of the overlap between eigenmodes and band-gaps of the associated infinite structures with different parameters, which in this particular case are the slit widths.

\begin{figure*}[htbp!]
    \centering
    \includegraphics[width=\linewidth]{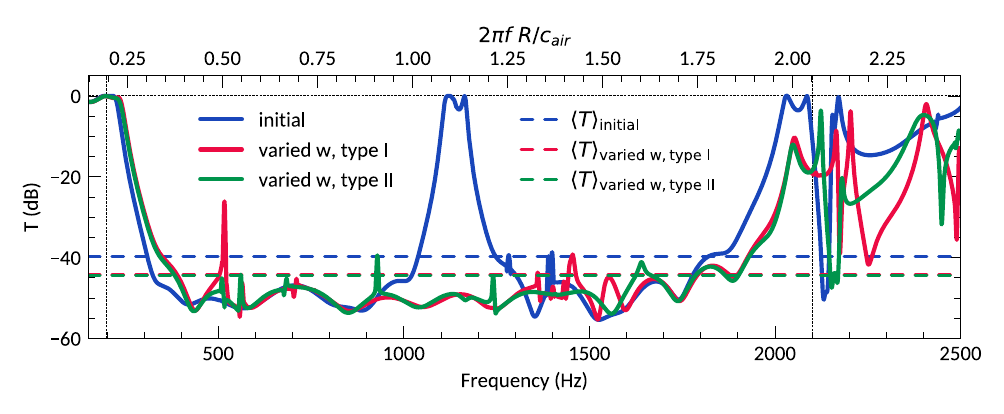}
    \caption{\textbf{Varied slit widths.} Transmission spectra for the initial case $w_{1-6} = 40$~mm and for the optimized structures with $w_1 = 36$, $w_2 = 62$, $w_3 = 46$, $w_4 = 77$, $w_5 = 103$, $w_6 = 68$~mm (type I) and $w_1 = 92$, $w_2 = 103$, $w_3 = 51$, $w_4 = 50$, $w_5 = 53$, $w_6 = 33$~mm (type II). Dashed horizontal lines indicate spectrally averaged transmission coefficient within the range $200$ -- $2100$~Hz.}
    \label{fig:GA_w_all_spectra}
\end{figure*}

In a more general case, when the resonators within the pair have different slit width, it is also possible to achieve similar results and obtain broad near-flat stop-band, as Fig.~\ref{fig:GA_w_all_spectra} shows. In this case the result was also obtained with the help of genetic algorithms.
The spectra in Fig.~\ref{fig:GA_w_all_spectra} looks quite similar to those presented in the main text. Nevertheless, the structures with non-equal slit widths within a unit cell are fundamentally different from the rest ones due to the break of the unit cell symmetry. While the associated effects are not studied, this is an additional degree of freedom which might be used for further tuning of the system.

\section{Lossy resonators}
\begin{figure}[htbp!]
    \centering
    \includegraphics[width=\linewidth]{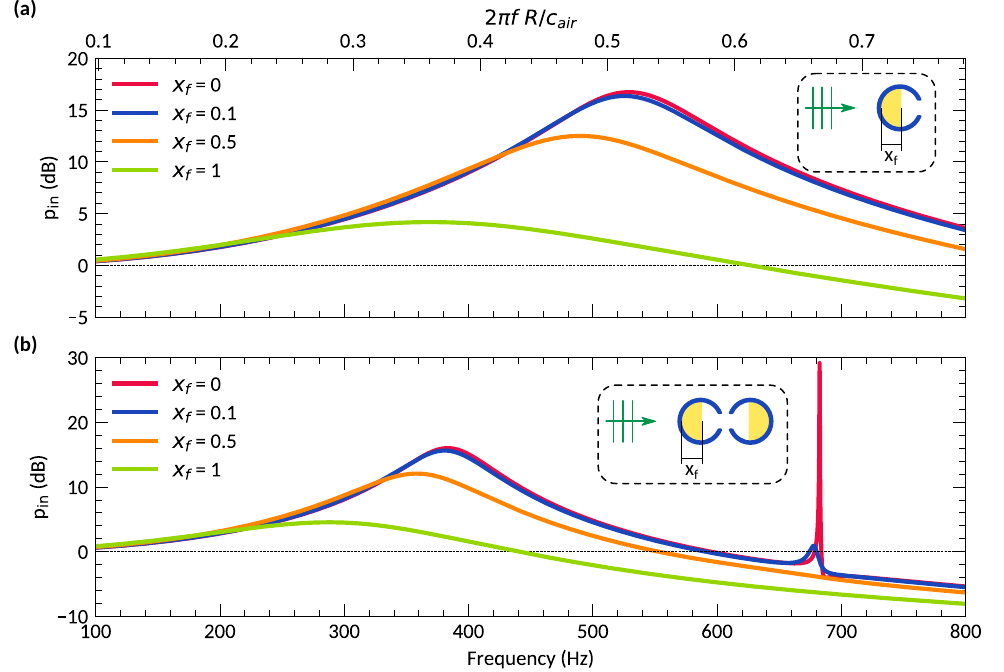}
    \caption{\textbf{Suppression of Helmholtz resonance in lossy resonators.} Pressure spectra inside (a) a single resonator and (b) the first resonator of a pair for different size of the porous insert $x_f$.}
    \label{fig:poroacoustics_Helmholtz}
\end{figure}

In the main text it is also stated that Helmholtz resonance is suppressed due to the losses related to poroacoustic inserts. This statement can be supported via consideration of a single and a pair of resonators in open space. 
Figure~\ref{fig:poroacoustics_Helmholtz} shows the corresponding spectra of pressure inside a resonator. In case of a single resonator [see Fig.~\ref{fig:poroacoustics_Helmholtz}(a)], increase of the poroacoustic insert size results in the decrease of Q-factor and shift of the resonance towards the lower frequencies. When the whole volume of the resonator is filled this practically results in the destruction of the resonance. Similarly, for the case of a pair of coupled resonators, both parts of the splitted resonance are fully suppressed when poroacoustic inserts occupy the whole volume of the resonators.
As in the main text, the flow resistivity of the insert is fixed at $6$~kPa$\cdot$s/m$^2$.

\begin{figure}[htbp!]
    \centering
    \includegraphics[width=\linewidth]{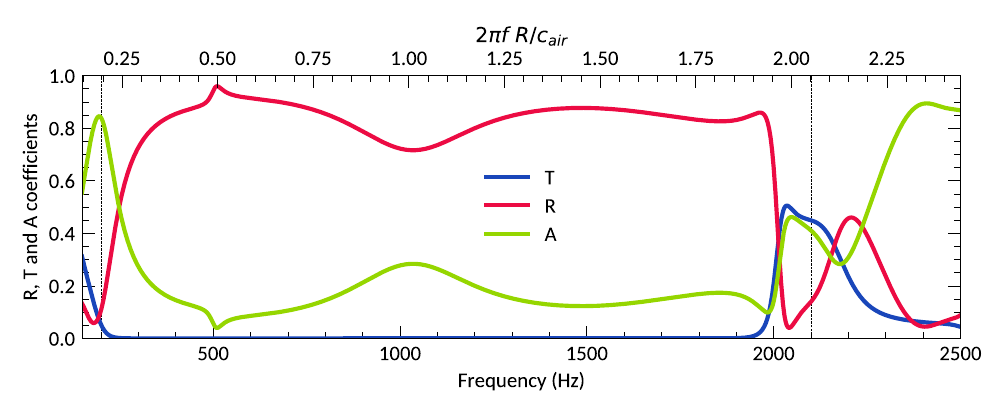}
    \caption{\textbf{Scattering and absorption of the periodic structure with poroacoustic inserts.} The spectra are caclulated for the case when the porous inserts occupy the whole volumes of the resonators.}
    \label{fig:poroacoustics_absorption_spectra}
\end{figure}

It also should be noted that the main mechanism of the noise-insulating properties is related not to absorption, but to periodicity of the considered structures. The absorption coefficient shown in Fig.~\ref{fig:poroacoustics_absorption_spectra} is rather low meaning that the most part of the energy is reflected by the structure. To explain this result, it can be speculated that the band-structure of the equivalent infinite system remains the same but porous inserts significantly reduce efficiency of the eigenmodes excitation. The absence of the transmission peak near $1100$~Hz in this case also support the statement that the corresponding eigenmode represent hybridized structural and Helmholtz resonances or collective Helmholtz resonance.

\begin{figure}[htbp!]
    \centering
    \includegraphics[width=\linewidth]{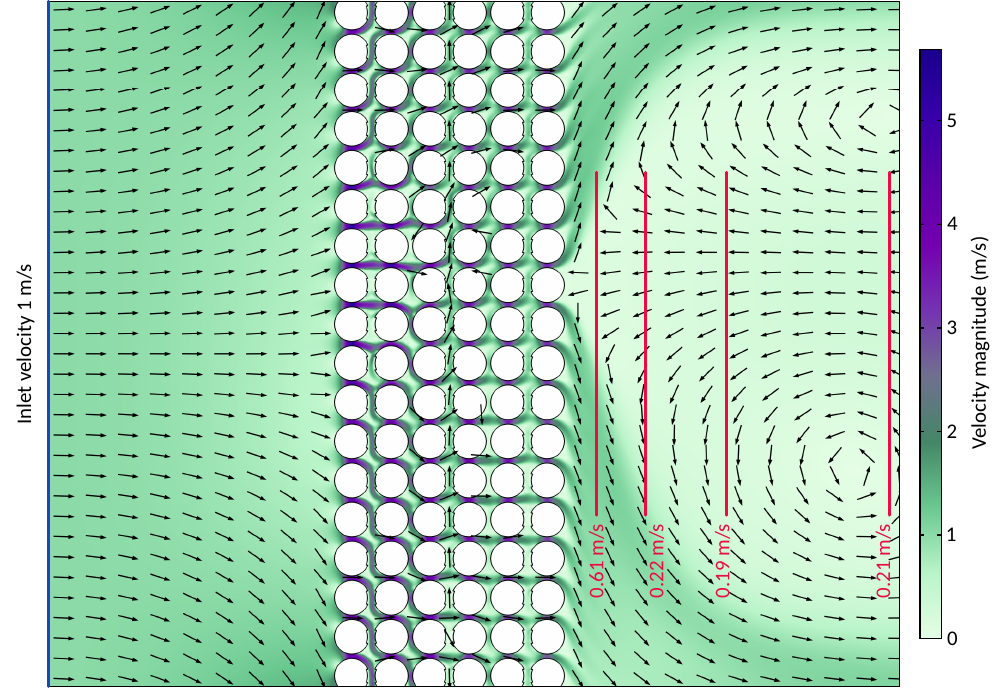}
    \caption{\textbf{Ventilation properties of the structure with lossy resonators.} The direction of the air flow propagation is shown by the black arrows and the magnitude of the flow velocity is indicated by color. The inlet velocity of the air flow is $1$~m/s. The magnitude of the velocity behind the structure is calculated as the value averaged over the red lines, located at the distance $0.1$, $0.25$, $0.5$ and $1$~m.}
    \label{fig:flow_wall_inserts}
\end{figure}

Despite rather good performance of the considered structure with the porous inserts, it might be expected that a porous layer of the same thickness might provide similar or even better results. However, a porous layer have a drawback of being non-ventilated contrary to the considered periodic structures. Indeed, Fig.~\ref{fig:flow_wall_inserts} shows that the air flow can propagate in-between the resonators and the flow velocity behind the structure is not zero. The flow distribution is rather complicated and asymmetric, which might be related to the non-trivial structure of the channels between the resonators. Nevertheless, the periodic structure do not fully block the air flow and a compromise between noise-insulation and ventilation properties can be found, which is not possible for a porous layer.
Calculations of the airflow through the structure are provided in Comsol Multiphysics, using the Turbulent Flow module. The RANS $k$--$\varepsilon$ turbulence model is used and the flow is considered to be weakly compressible. The inlet wall is characterized by the normal inlet velocity $1$~m/s. The backflow at the outlet wall is suppressed and the pressure boundary condition is used, such that the pressure is equal to $0$~Pa at the outlet wall boundary.

\end{document}